\documentclass[letterpaper, 10 pt, conference]{ieeeconf}  
\IEEEoverridecommandlockouts                              

\usepackage{amsmath} 
\usepackage{amssymb}  
\usepackage{graphicx}

\usepackage{subcaption}
\usepackage{mathtools}
\usepackage{commath}

\usepackage{bbm}

\usepackage[]{algorithm2e}
\usepackage{algorithmic}
\SetAlgoCaptionSeparator{.}
\usepackage{caption}
\usepackage[table]{xcolor}
\usepackage{multirow}
\usepackage{array}
\usepackage{float}

\title{\LARGE \bf
Trajectory planning with a dynamic obstacle clustering strategy using Mixed-Integer Linear Programming*
}

\author{Vin\'{i}cius Antonio Battagello$^{1}$, Nei Yoshihiro Soma$^{1}$ and Rubens Junqueira Magalh\~{a}es Afonso$^{2,3}$
\thanks{*This work was supported by %
					Conselho Nacional de Desenvolvimento Cient\'{i}fico e Tecnol\'{o}gico CNPq (grant 141007/2016-8), S\~{a}o Paulo Research Foundation - FAPESP (grant 2016/01860-1), CNPq PQ, and %
					CAPES (grant 88881.145490/2017-01) and the German Ministry of Education and Research through the Alexander von Humboldt Foundation %
					}
\thanks{$^{1}$Vin\'{i}cius A. Battagello is with the Divis\~{a}o de Engenharia de Computa\c{c}\~{a}o, Instituto Tecnol\'{o}gico de Aeron\'{a}utica, S\~{a}o Jos\'{e} dos Campos, SP 12228-900, Brazil %
        {\tt\small battagello@ita.br}}%
\thanks{$^{2}$Nei Y. Soma is with the Divis\~{a}o de Engenharia de Computa\c{c}\~{a}o, Instituto Tecnol\'{o}gico de Aeron\'{a}utica, S\~{a}o Jos\'{e} dos Campos, SP 12228-900, Brazil %
        {\tt\small soma@ita.br}}%
\thanks{$^{3}$Rubens J. M. Afonso is with the Institute of Flight System Dynamics, Department of Aerospace and Geodesy, Technical University of Munich (TUM), Garching bei M\"{u}nchen, Bavaria, Germany %
        {\tt\small rubens.afonso@tum.de}}%
\thanks{$^{4}$Rubens J. M. Afonso is on leave from the Divis\~{a}o de Engenharia Eletr\^{o}nica, Instituto Tecnol\'{o}gico de Aeron\'{a}utica, S\~{a}o Jos\'{e} dos Campos, SP 12228-900, Brazil %
        {\tt\small rubensjm@ita.br}}%
}
\begin{document}

\maketitle
\thispagestyle{empty}
\pagestyle{empty}

\begin{abstract}

In this paper we propose a technique that assigns obstacles to clusters used for collision avoidance via Mixed-Integer Programming. This strategy enables a reduction in the number of binary variables used for collision avoidance, thus entailing a decrease in computational cost, which has been a hindrance to the application of Model Predictive Control approaches with Mixed-Integer Programming formulations in real-time. Moreover, the assignment of obstacles to clusters and the sizes of the clusters are decided within the same optimization problem that performs the trajectory planning, thus yielding optimal cluster choices. Simulation results are presented to illustrate an application of the proposal.

\end{abstract}


\section{INTRODUCTION}


Motion planning in the presence of obstacles is of paramount importance for autonomous agents. 
Generally, motion planning may be split into three different layers, namely (i) path planning, (ii) trajectory planning, and (iii) low-level control. 

Path planning methods aim at finding routes in the euclidean space without time parametrization, whereas trajectory planning deals with the search for routes that obey dynamics constraints, usually assigning a time tag to each position in space.

Some proposals combine both path and trajectory planning in a single layer with advantages regarding the optimal cost of the ensuing trajectory. 
Current successful approaches involve using Model Predictive Control (MPC) formulations based on the solution of a Mixed-Integer Linear Program (MILP) to cope with the nonconvexity brought about by the presence of obstacles. 
In \cite{Richards2002} the authors propose an approximate model of aircraft dynamics using linear constraints to the trajectory planning of airplanes with the application of a MILP approach. The formulation ensures collision avoidance for each aircraft and guarantees the desired hard constraints fulfillment. This approach is extended in \cite{Richards2006}, which applies a constraint tightening strategy to obtain a robust solution that guarantees finite-time arrival into an arbitrary target set and fulfillment of output/input constraints, in spite of unknown disturbance. The central idea is to  hold a ``border'' for feedback action as time goes by.

However, MILP is known to be NP-hard \cite{Garey1979}, thus such approaches may demand a prohibitive computational effort for complex environments. The computation time demanded in trajectory planning grows strongly with the number of binary variables used for obstacle avoidance, therefore this has been the aim of several recent contributions both using pre-processing and alternative MILP encodings.  

In \cite{Vitus2008} a three-stage algorithm is used to perform pre-processing: (i) a collision-free path is computed via graph search, (ii) then convex polytopes that cover the entire path are generated by using a triangulation of the free space, and finally (iii) a MILP formulation using the polytopes as a ``tunnel'' from the starting position to the goal is solved to determine the optimal trajectory. Since it relies on a graph search to prune the free space in the first step, optimality of the planned trajectory cannot be guaranteed, as the cost is not taken into account in this initial phase.

Another example of pre-processing is proposed in \cite{Deits2015}, where ellipsoidal regions of convex obstacle-free space are calculated to provide a collision-free path with fewer integer variables. 

More recently, several techniques using both pre-processing of the environment as well as taking advantage of the current available solution of the MPC problem were proposed in \cite{Battagello2020}. Regarding pre-processing, the proposal involved clustering the obstacles according to their distances to the agent and among themselves, replacing several obstacles and the binary variables associated to them with fewer clusters and their corresponding binary variables. Moreover, as the MPC is run in a receding horizon fashion, the current available solution was explored to determine which obstacles can be replaced by convex constraints as they are not circled by the agent.

On the other hand, in \cite{Prodan2012} an alternative encoding of the exclusion regions in terms of the binary variables associated to them is proposed, leading to a decrease in the number of binary variables. In fact, a linear dependency of the number of binary variables with the number of sides of the obstacles was reduced to a logarithmic growth. The number of additional inequalities to allow that was shown to be one per obstacle in \cite{Afonso2013}. For a more thorough discussion of the MILP encoding schemes used in control the reader is referred to \cite{Prodan2015}.

In the present paper, we propose a novel scheme to perform the clustering of the obstacles within the same optimization problem that is solved to yield the optimal trajectory. Therefore, a reduction in the online computational time is provided, whereas the clustering is directly coupled with the cost optimization, these two characteristics being the main advantages of our proposal.

The remainder of this paper is divided as follows: the problem is formally presented in Sec. \ref{sec:2-problem_statement} along with a detailed description of how the obstacle avoidance is performed in a MILP framework; the proposed clustering strategy and the constraints to avoid collisions with the clusters are developed in Sec. \ref{sec:2-cluster_restrainment}; the simulation scenario is presented in Sec. \ref{sec:2-simulation_scenarios}; Sec. \ref{sec:3-results} contains the simulation results and their discussion; concluding remarks and suggestions for future work are given in Sec. \ref{sec:3-conclusions}.

\section{PROBLEM STATEMENT}

\label{sec:2-problem_statement}

Here we study the problem of maneuvering an agent in a plane with a coordinate system $(r_x,r_y)$
with state $\textbf{x} \in \mathbb{R}^4$ and position $\textbf{r} = [ r_x \ r_y ]^T \in \mathbb{R}^2$  into a target region in a two-dimensional environment, while collisions against obstacles must be avoided. Extending the proposal to a three-dimensional scenario is straightforward and is not treated in the present paper for simplicity of presentation.

The dynamics of the agent in discrete-time is given as:
\begin{equation}
		\textbf{x}[k+1] = \textbf{A} \textbf{x}[k] + \textbf{B} \textbf{u}[k].				\label{eq:system_dynamics} 
\end{equation}

MPC is used to perform the maneuver of the system in \eqref{eq:system_dynamics}. The task of the control problem is to choose optimally the predictions	$\hat{\textbf{x}}[k+j \vert k]$ with the corresponding $\hat{\textbf{u}}[k+j \vert k]$ in each timestep $j$. The maneuvering is performed in the literature \cite{Richards2002, Bellingham2002, Richards2006} by
solving the following optimization problem:

\begin{equation}			\label{eq:cost_function0}
	J \left[ \textbf{x}[k] \right] = 	\min_{\hat{ \textbf{u}}, \hat{\textbf{x}}, N[k]} \sum_{j=0}^{N[k]} \left(1 + \gamma \norm{ \hat{\textbf{u}} [k+j \vert k]}_1 \right)
\end{equation}
subject to

\begin{subequations}
	\begin{flalign}						
			& \hat{ \textbf{x}} [k|k] = \textbf{x}[k],									\label{eq:x_current_state_known} 		\\
			& \hat{ \textbf{x}} [k+j \vert k] \in \mathcal{X} , \ j = 1,2, \ldots, N[k]+1 \label{eq:x_set} \\
			& \hat{ \textbf{u}} [k+j \vert k] \in \mathcal{U} , \ j = 0, 1, \ldots, N[k]		\label{eq:u_set}	\\
			& \hat{ \textbf{r}} [k+j \vert k] = \textbf{C} \hat{ \textbf{x} }[k+j|k],	\ j = 1, 2, \ldots, N[k]+1	\label{eq:r_extracted_from_x} 	\\
			& \hat{ \textbf{r}} [k+j \vert k] \notin \mathcal{O}_i	 ,	\ j = 1, 2, \ldots, N[k]+1, \ i = 1,2,\ldots,N_o \label{eq:r_set} 		\\
			& \hat{ \textbf{x}} [k+N[k]+1 \vert k] \in \mathcal{Q} , \label{eq:terminal_set} 									\\		
			& \hat{ \textbf{x}} [k+j+1 \vert k] \ = \ \textbf{A} \hat{ \textbf{x}} [k+j \vert k]+ \textbf{B} \hat{\textbf{u}}[k+j \vert k] ,		\label{eq:system_dynamics_estimate}  \\
			& j = 0, 1, \ldots, N[k], 							\label{eq:j_set}
	\end{flalign}
\end{subequations}
\noindent where $N[k]$ is a variable horizon, $\mathcal{X} \subset \mathbb{R}^4$ is a polytope representing the pointwise-in-time state constraints, $\mathcal{U} \subset \mathbb{R}^2$ is a polytope representing the pointwise-in-time control constraints, $\mathcal{Q} \subset \mathbb{R}^4$ is a polytope representing the terminal constraints,
and $\textbf{C}  \in \mathbb{R}^{2 \times 4}$ is a matrix that extracts position information from the state vector. The sets $\mathcal{O}_i \subset \mathbb{R}^2$, $i = 1,2,\ldots, N_o$, represent the $N_o$ polytopic obstacles that must be avoided.

For details on how to rewrite this optimization problem as a MILP with a finite horizon $N_s$ the reader is referred to \cite{Richards2002, Bellingham2002, Richards2006}. 
In the present paper we present only the obstacle avoidance constraints in detail as they are the object of our contribution.

\subsection{Obstacle avoidance}		\label{sec:2-obstacle_avoidance_constraints} 
\label{chap:unclustered_strategy}
The presence of obstacles may render the problem of optimizing a trajectory not convex. In this section we describe the commonly adopted remodeling used in the MPC-MILP literature with binary optimization variables and the ``big-M'' approach \cite{Bemporad1999}. 

For the sake of simplicity, the obstacles are assumed to be rectangles with sides parallel to the coordinate axes. Generalizing to polytopes of any number of sides is straightforward and our simplification does not represent a limitation of the proposal. Bearing that in mind, the obstacles $\mathcal{O}_i, \, i = 1, 2,\dots, \, N_o$ can be represented by:
\begin{equation}
	\mathcal{O}_i = \lbrace \textbf{z} \in \mathbb{R}^2 | [o_{i}^{x,min}  \ o_{i}^{y,min}]^T \leq \textbf{z} \leq [o_{i}^{x,max} \ o_{i}^{y,max}]^T] \rbrace,
\end{equation}
\noindent where 
$o_{i}^{x,min}$, $o_{i}^{y,min}$, $o_{i}^{x,max}$, and $o_{i}^{y,max}$ represent the coordinates of the left, lower, right, and upper sides of the i-th obstacle, respectively.

Collisions can be avoided by imposing that the position of the agent is in one of the outer half-planes defined by the support lines of the sides of the obstacle. This is done by using the binary variable $b_{j,i,f}^{\mathcal{O}}$, which marks at time step $k+j$ if the agent is outside side $f$ of obstacle $\mathcal{O}_i$. Such constraints may be written using the ``big-M'' method as
\begin{subequations}
	\begin{alignat}{1}
		 \hat{r}_x[k+j|k] &\leq  \, \ \ o_{i}^{x,min} + b_{j,i,1}^{\mathcal{O}} M_\mathcal{O},	\label{eq:obst_avoid_constraints_left}	\\
		 \hat{r}_y[k+j|k] &\leq  \, \ \ o_{i}^{y,min} + b_{j,i,2}^{\mathcal{O}} M_\mathcal{O},	\label{eq:obst_avoid_constraints_down}	\\
		-\hat{r}_x[k+j|k] &\leq - o_{i}^{x,max} + b_{j,i,3}^{\mathcal{O}} M_\mathcal{O},	\label{eq:obst_avoid_constraints_right}		\\
		-\hat{r}_y[k+j|k] &\leq - o_{i}^{y,max} + b_{j,i,4}^{\mathcal{O}} M_\mathcal{O},	\label{eq:obst_avoid_constraints_up}		\\
		 \sum_{f=1}^{4} b_{j,i,f}^{\mathcal{O}}  &\leq 3, \ j = 1,2,\ldots,N_s, \ i = 1,2,\ldots, N_o, \label{eq:obst_avoid_constraints_bin}
	\end{alignat}			\label{eq:obst_avoid_constraints}
\end{subequations} 

\noindent where $M_\mathcal{O} \in \mathbb{R}^+$ is chosen large enough to render the inequalities \eqref{eq:obst_avoid_constraints_left}, \eqref{eq:obst_avoid_constraints_down}, \eqref{eq:obst_avoid_constraints_right} and \eqref{eq:obst_avoid_constraints_up} inactive for all achievable values of $\hat{r}_x[k+j|k]$ and  $\hat{r}_y[k+j|k]$, according to the ``big-M'' method. 

If we use this scheme to model environments with increasingly more obstacles, as each obstacle partitions the search space into four disjoint regions, the computation time in environments with tens of obstacles or more becomes prohibitive. Therefore, in this paper we propose to group the obstacles in clusters and perform the collision avoidance with the clusters instead of the obstacles, \emph{i.e.} using binary variables $b_{j,\ell,f}^{\mathcal{C}}$ for cluster avoidance rather than $b_{j,i,f}^{\mathcal{O}}$ for obstacle avoidance.

%
%

\section{PROPOSED CLUSTER-OBSTACLE ASSIGNMENT AND COLLISION AVOIDANCE} 		\label{sec:2-cluster_restrainment}

In this section we will present our proposal to enable avoiding clusters of obstacles, with the difference regarding the literature that the definition of the clusters is also an object of optimization within the MILP problem.

Let the $N_c$ clusters be represented as $\mathcal{C}_\ell \subset \mathbb{R}^2$, $\ell=1,2,\ldots,N_c$. As in section \ref{chap:unclustered_strategy}, we consider rectangular clusters with sides parallel to the coordinate axes for simplicity. Again, this simplification serves the purpose of ease of presentation, but represents no limitation of the proposal, as it can seamlessly be adapted for general polytopic sets. The first requirement is that each obstacle $\mathcal{O}_i$ must be a subset of at least one of the clusters $\mathcal{C}_\ell$, so that avoiding collisions with the clusters entails avoiding collisions with the obstacles. This is formally stated as the ``\textbf{or}'' constraints:
\begin{equation}		\label{eq:cluster_restraintment_definition}
		\mathcal{O}_i \subset \mathcal{C}_1 \lor \mathcal{O}_i \subset \mathcal{C}_2 \lor \ldots \lor \mathcal{O}_i \subset \mathcal{C}_{N_c}, \ i = 1,2,\ldots,N_o.
\end{equation}

It is easy to see that, if \eqref{eq:cluster_restraintment_definition} holds, then avoiding collisions with the obstacles as required by \eqref{eq:r_set} can be done by avoiding collisions with the clusters, since
\begin{equation}
		\hat{ \textbf{r}} \notin \mathcal{C}_\ell, \ \ell = 1,2,\ldots,N_c \implies \hat{ \textbf{r}} \notin \mathcal{O}_i, \ i = 1,2,\ldots,N_o. \label{eq:cluster_implies_obstacle}
\end{equation}

Since the constraints in \eqref{eq:cluster_restraintment_definition} are of the logical type ``\textbf{or}'', again binary variables and the ``big-M'' are one tool to encode them within the MILP formulation \cite{Bemporad1999}. Then, using binary variables $b^R_{\ell,i}$ we can implement \eqref{eq:cluster_restraintment_definition} through the following inequalities:
\begin{subequations}
	\begin{align}
			c^{x,min}_\ell 		&\leq  \, \ \ o_{i}^{x,min} + (1-b^R_{\ell,i})M_c, 	\label{eq:cluster_restrainment_constraints_left}			\\
			c^{y,min}_\ell 		&\leq  \, \ \ o_{i}^{y,min} + (1-b^R_{\ell,i})M_c,		\label{eq:cluster_restrainment_constraints_down}	\\
			-c^{x,max}_\ell 		&\leq -o_{i}^{x,max} + (1-b^R_{\ell,i})M_c,			\label{eq:cluster_restrainment_constraints_right}	\\
			-c^{y,max}_\ell 		&\leq -o_{i}^{y,max} + (1-b^R_{\ell,i})M_c ,			\label{eq:cluster_restrainment_constraints_up}	\\
			\sum_{\ell = 1}^{N_c} b^R_{\ell,i} & = 1, \ i = 1,2,\ldots,N_o, \label{eq:cluster_restrainment_constraints_bin}
\end{align}			\label{eq:cluster_restrainment_constraints}
\end{subequations} 

\noindent where $c^{x,min}_\ell$, $c^{y,min}_\ell$, $c^{x,max}_\ell$, and $c^{y,max}$ are decision variables that represent the coordinates of the left, lower, right, and upper sides of cluster $\mathcal{C}_\ell$, $\ell = 1,2,\ldots, N_c$, respectively. Constraint \eqref{eq:cluster_restrainment_constraints_bin} ensures that for each obstacle $\mathcal{O}_i$, at least one of the binary variables $b^R_{\ell,i}$ is set to one, thus assigning it to the corresponding cluster $\mathcal{C}_\ell$.

\begin{figure}[bp]
      \centering							
						\includegraphics[width=0.7\linewidth]{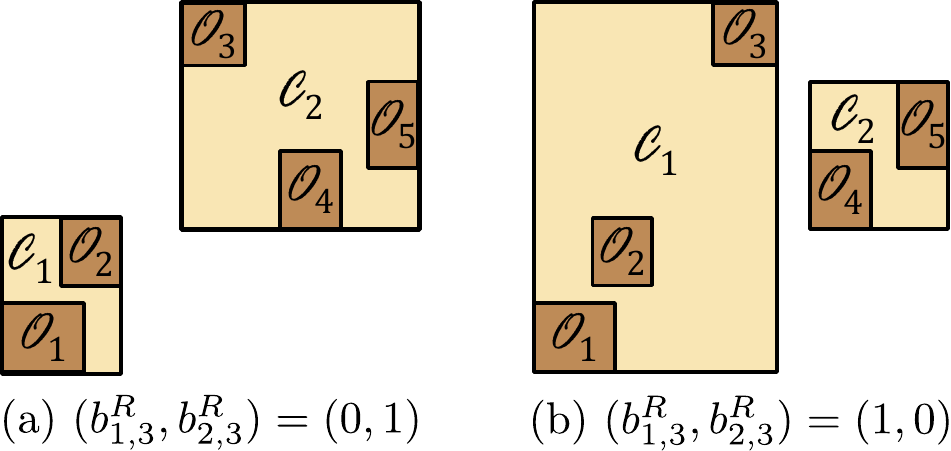}						
      \caption{Examples of cluster-obstacle assignment for $N_c = 2$ and $N_o = 5$.}
      \label{fig:cluster_restrainment_concept}
   \end{figure}

Fig.~\ref{fig:cluster_restrainment_concept} depicts examples with the same obstacles $\mathcal{O}_i$, $i = 1,2,\ldots, 5$ and two clusters $\mathcal{C}_1$ and $\mathcal{C}_2$. The corresponding values of the cluster-obstacle assignment binary variables $b^R_{\ell,i}$ are shown in Tab. \ref{tab:ClusterAssignmentExample} for the cases ``a'' and ``b'' depicted in Fig.~\ref{fig:cluster_restrainment_concept}a and Fig.~\ref{fig:cluster_restrainment_concept}b, respectively. The difference in the clustering choices seen in Fig.~\ref{fig:cluster_restrainment_concept} can be explained by the changes in the value of the binary variables $b_{1,3}^R$ and $b_{2,3}^R$ between cases ``a'' and ``b'' in Tab. \ref{tab:ClusterAssignmentExample}. In case ``a'', $b_{2,3}^R=1$, effectively imposing that $\mathcal{O}_3 \subset \mathcal{C}_2$. On the other hand, in case ``b'', $b_{1,3}^R=1$, thus entailing $\mathcal{O}_3 \subset \mathcal{C}_1$.

\begin{table*}[htbp]
\vspace{0.3cm}
	\centering
		\begin{tabular}{r|cccccccccc}
		 & \multicolumn{10}{c}{Binary variables} \\
			Case & $b^R_{1,1}$ & $b^R_{1,2}$ & $b^R_{1,3}$ & $b^R_{1,4}$ & $b^R_{1,5}$ & $b^R_{2,1}$ & $b^R_{2,2}$ & $b^R_{2,3}$ & $b^R_{2,4}$ & $b^R_{2,5}$ \\
			a & 1 & 1 & 0 & 0 & 0 & 0 & 0 & 1 & 1 & 1 \\
			b & 1 & 1 & 1 & 0 & 0 & 0 & 0 & 0 & 1 & 1
		\end{tabular}
	\caption{Cluster-obstacle assignment binary variables $b^R_{\ell,i}$ for the example in Fig. \ref{fig:cluster_restrainment_concept}.}
	\label{tab:ClusterAssignmentExample}
\end{table*}

Cluster avoidance can then be ensured by imposing that $\textbf{r} \notin \mathcal{C}_\ell, \ell = 1,2,\ldots, N_c$, which can be written in MILP form using cluster avoidance binary variables $b_{j,\ell,f}^{\mathcal{C}}$ and the ``big-M'' method as
\begin{subequations}
	\begin{alignat}{1}
		 \hat{r}_x[k+j|k] &\leq  \, \ \ c^{x,min}_\ell + b_{j,\ell,1}^{\mathcal{C}} M_\mathcal{C}		\label{eq:cluster_avoidance_constraints_left} \\
		 \hat{r}_y[k+j|k] &\leq  \, \ \ c^{y,min}_\ell + b_{j,\ell,2}^{\mathcal{C}} M_\mathcal{C}		\label{eq:cluster_avoidance_constraints_down} \\
		-\hat{r}_x[k+j|k] &\leq - c^{x,max}_\ell + b_{j,\ell,3}^{\mathcal{C}} M_\mathcal{C}		\label{eq:cluster_avoidance_constraints_right}	\\
		-\hat{r}_y[k+j|k] &\leq - c^{y,max}_\ell + b_{j,\ell,4}^{\mathcal{C}} M_\mathcal{C}		\label{eq:cluster_avoidance_constraints_up}	\\
		 \sum_{f=1}^{4} b_{j,\ell,f}^{\mathcal{C}}  &\leq 3, \ j = 1,2,\ldots,N_s, \ \ell = 1,2,\ldots, N_c, \label{eq:cluster_avoidance_constraints_bin}
	\end{alignat}			\label{eq:cluster_avoidance_constraints}
\end{subequations} 

\noindent where $M_\mathcal{C} \in \mathbb{R}^+$ is chosen large enough to render the inequalities \eqref{eq:cluster_avoidance_constraints_left}, \eqref{eq:cluster_avoidance_constraints_down}, \eqref{eq:cluster_avoidance_constraints_right} and \eqref{eq:cluster_avoidance_constraints_up} inactive for all achievable values of $\hat{r}_x[k+j|k]$ and  $\hat{r}_y[k+j|k]$, according to the ``big-M'' method. 

Whereas usually the constraints in \eqref{eq:obst_avoid_constraints} are employed in the literature to implement \eqref{eq:r_set}, requiring $4 N_s N_o$ binary variables $b^\mathcal{O}_{j,i,f}$, in our proposal \eqref{eq:cluster_avoidance_constraints} is used to impose $\hat{\textbf{r}}[k+j \vert k] \notin \mathcal{C}_\ell$, requiring $4 N_s N_c$ binary variables $b^\mathcal{C}_{j,\ell,f}$. Moreover, in view of \eqref{eq:cluster_implies_obstacle}, one must append constraints \eqref{eq:cluster_restrainment_constraints} to the problem in our proposal to ensure that \eqref{eq:cluster_restraintment_definition} holds, which in turn demands $N_o N_c$ additional binary variables $b^R_{\ell,i}$. Therefore, the total number of of binary variables to avoid collisions with our proposal amounts to $4 N_s N_c + N_o N_c = (4 N_s + N_o) N_c$. Notice that in the state-of-the-art the product between the number of obstacles $N_o$ and the maximal horizon $N_s$ dictates the growth in the number of binary variables. In contrast, with our proposal, this product does not appear and both $N_s$ and $N_o$ are multiplied by the number of clusters $N_c$. Since the number of obstacles and the maximal horizon are linked to the scenario itself, they cannot be controlled by the user to achieve a lower computational burden. In opposition, with our approach the user can decide the number $N_c$ and effectively limit the growth of the number of binary variables. 

Clearly, the lowest $N_c$, the farthest from the optimal cost the solution yielded with our proposal, as the possibility of passing between many obstacles is hindered by clustering them. Therefore, one expects a compromise between the computational workload and the optimality of the planned trajectory. However, it is important to emphasize that although our proposal entails this compromise, one key advantage is that the clusters are decided within the same optimization problem that minimizes the cost, as opposed to techniques that try to eliminate the obstacles or to produce a collision-free tunnel before calculating the optimal trajectory, as mentioned in the literature review. Furthermore, it is interesting to remark that the controller runs in a receding horizon fashion, therefore the optimization problem is solved at each timestep with feedback from the current measured state. As a consequence, the clustering can be changed throughout the maneuver.


\vspace{-0.2cm}
\section{SIMULATION SCENARIO} 		\label{sec:2-simulation_scenarios}

We consider in this work the model of an agent as that of a particle moving in a plane with axes $r_x$ and $r_y$ orthogonal to each other, as in \cite{Richards2002, Bellingham2002, Richards2006}. The position vector is defined as $\textbf{r} = [r_x \ r_y]^T$, the inputs are the accelerations $a_x$ and $a_y$, which yield velocities $v_x$ and $v_y$ aligned respectively with the $r_x$ and $r_y$ axes.
The continuous-time model in state-space is given by $\dot{\textbf{x}} = \textbf{A}_\textbf{c} \textbf{x} + \textbf{B}_\textbf{c} \textbf{u}$, with the state $\textbf{x}$ and control $\textbf{u}$ vectors given by
$\textbf{x}^T = [r_x \ v_x \ r_y \ v_y]$ and $\mathbf{u}^T = [a_x \ a_y]$. 


Since the MPC controller is implemented in discrete-time, we must discretize the plant model to use it as an internal controller, which is done by Zero-Order Hold discretization \cite{Franklin1998}.
This generates a model in the form ${\textbf{x}[k+1] = \textbf{A} \textbf{x}[k] + \textbf{B} \textbf{u}[k]}$, and by choosing a sample period of $0.8$ time units, matrices $\textbf{A}$ and $\textbf{B}$ are given in \eqref{eq:ab_discrete_matrix}. Here, units of length and time are assumed to be $m$ and $s$. 
 	\begin{minipage}{.25\textwidth}\ttfamily
			\begin{equation*}
					\mathbf{A} = \begin{bmatrix}
    		 			1 &  0.8 &   0 &  0 \\ 
        	 			0 &  1 &  0 &   0 \\
        	 		    0 &  0 &  1 &  0.8  \\
        	 			0 &  0 &  0 &  1  
    				\end{bmatrix}  ,
		\end{equation*}
	\end{minipage}%
	\begin{minipage}{.25\textwidth}\ttfamily
		\begin{equation}
			\mathbf{B} = \begin{bmatrix}
    		 0.32 &  0 \\ 
        	 0.8 &  0 \\
         	 0 &  0.32 \\
            0 &  0.8
	    	\end{bmatrix}  .
    		\label{eq:ab_discrete_matrix}
		\end{equation}		
	\end{minipage}

\vspace{0.2cm}

The controller parameters are
\begin{itemize}
	\item the weight of the sum of the absolute values of the accelerations $\gamma = 1$,
	\item the maximal horizon $N_{s} = 18$,
	\item the constraints over velocities and accelerations at every time step: $-10 \ m/s \leq v_x, v_y \leq 10 \ m/s$ and $-3 \ m/s^2 \leq a_x, a_y \leq 3 \ m/s^2$,
	\item and the terminal set $\mathcal{Q} = \left\{ \textbf{x} \in \mathbb{R}^4 \vert  
 [14 \ -v_{\epsilon} \ 0 \ -v_{\epsilon}]^T \leq \textbf{x} \leq [15 \ v_{\epsilon} \ 1 \ v_{\epsilon}]^T\right\}$, for $v_{\epsilon} = 5 \times 10^{-3} \ m/s$.
\end{itemize}

The initial state is $\textbf{x}[0] = [0 \ 0 \ 10 \ 0]^T$and the scenario has $N_c = 45$ obstacles, which are depicted as brown rectangles in the plots.

The CPLEX toolbox from IBM ILOG was used for solving the MILP problem in Matlab environment.

\section{RESULTS AND DISCUSSION}		\label{sec:3-results}
Three different simulations are run for comparison: (i) state-of-the-art collision avoidance employing the constraints in Sec. \ref{sec:2-obstacle_avoidance_constraints}, herein deemed ``unclustered'', collision avoidance using the proposal in Sec. \ref{sec:2-cluster_restrainment} with (ii) $N_c = 2$ and (iii) $N_c = 3$ clusters, herein deemed ``clustering''.

\begin{figure}[thpb]
      \centering		
						\includegraphics[width=0.7\linewidth]{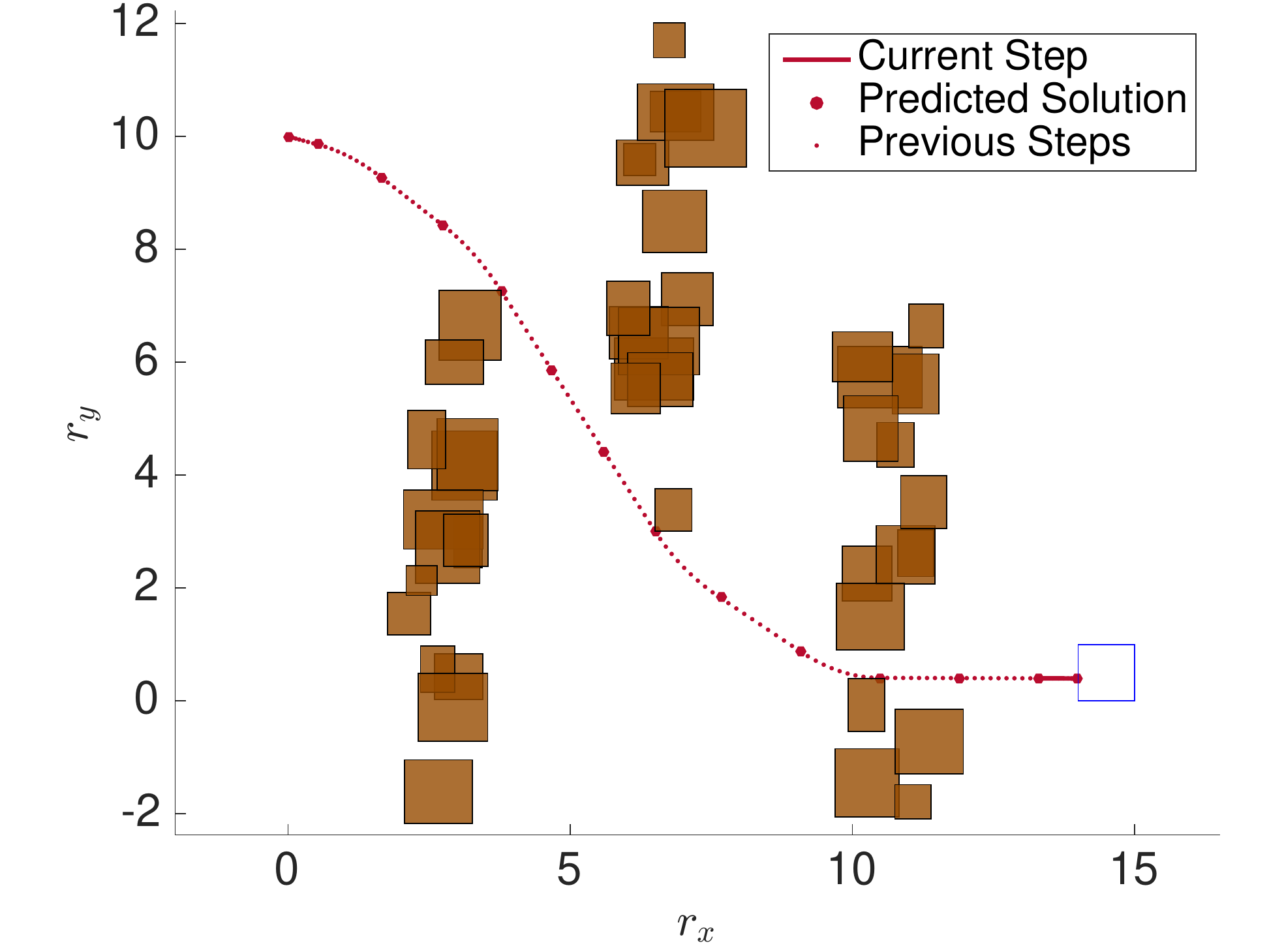}						
      \caption{Trajectory of the Unclustered avoidance strategy.}
      \label{fig:trajectory_unclustered}
   \end{figure}

\begin{figure}[thpb]
      \centering		
						\includegraphics[width=0.7\linewidth]{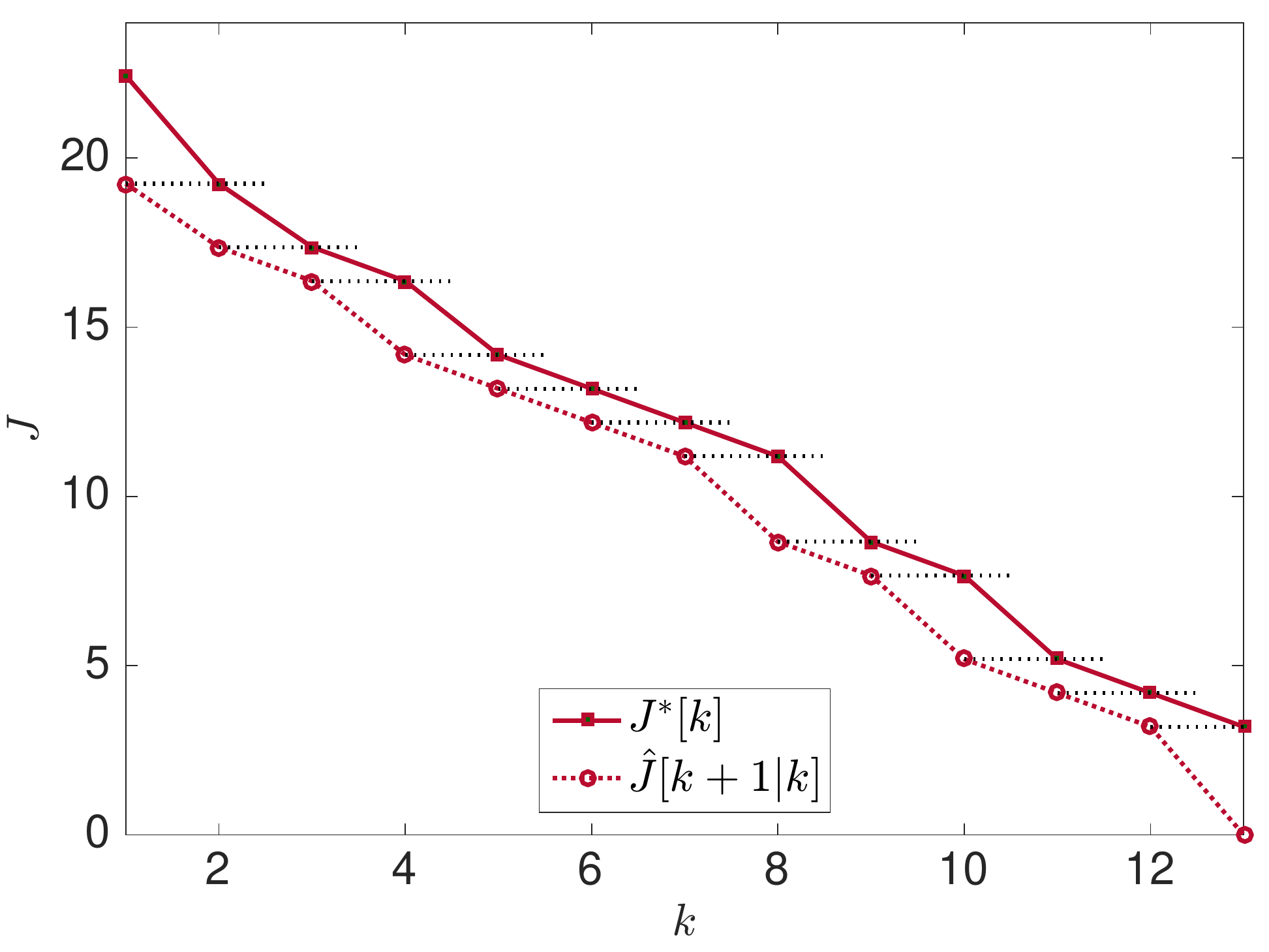}					
      \caption{Evolution of the cost function and its prediction for the Unclustered avoidance strategy.}
      \label{fig:Jstar_Jhat_unclustered}
   \end{figure}

The results of the execution of the unclustered collision avoidance strategy can be found in the trajectory of Fig.~\ref{fig:trajectory_unclustered}, with the evolution of the cost function $J^\star[k]$ and its prediction $\hat{J}[k+1|k]$ depicted in Fig.~\ref{fig:Jstar_Jhat_unclustered}.

\begin{figure}[htbp]
\centering
	\begin{subfigure}{0.35\textwidth}																																							
		\includegraphics[width=1.00\linewidth]{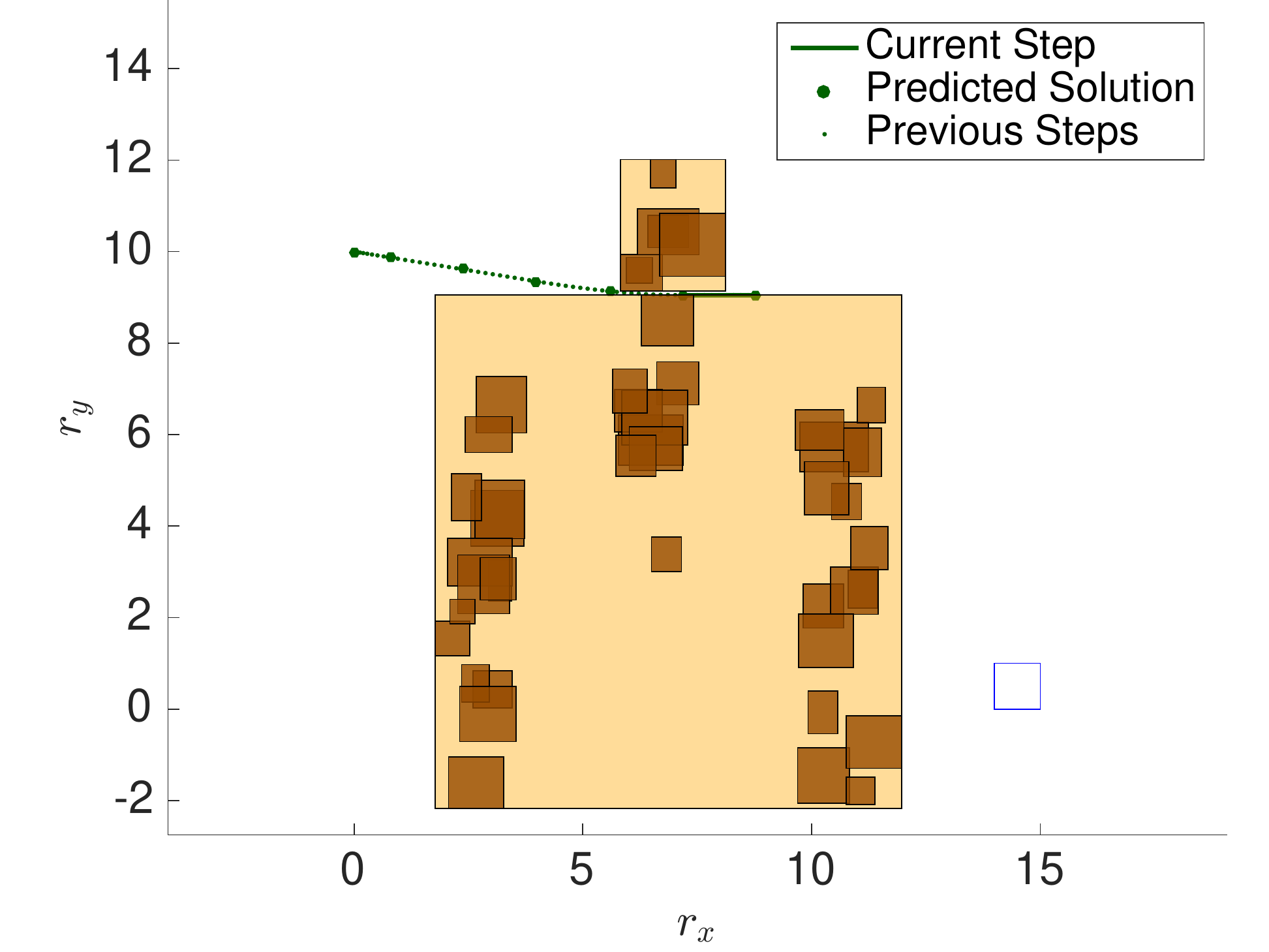} 
		\caption{$k = 6$}
		\label{fig:MILPclust_Nc2_step6}
	\end{subfigure}
	
	\begin{subfigure}{0.35\textwidth}																																							
		\includegraphics[width=1.00\linewidth]{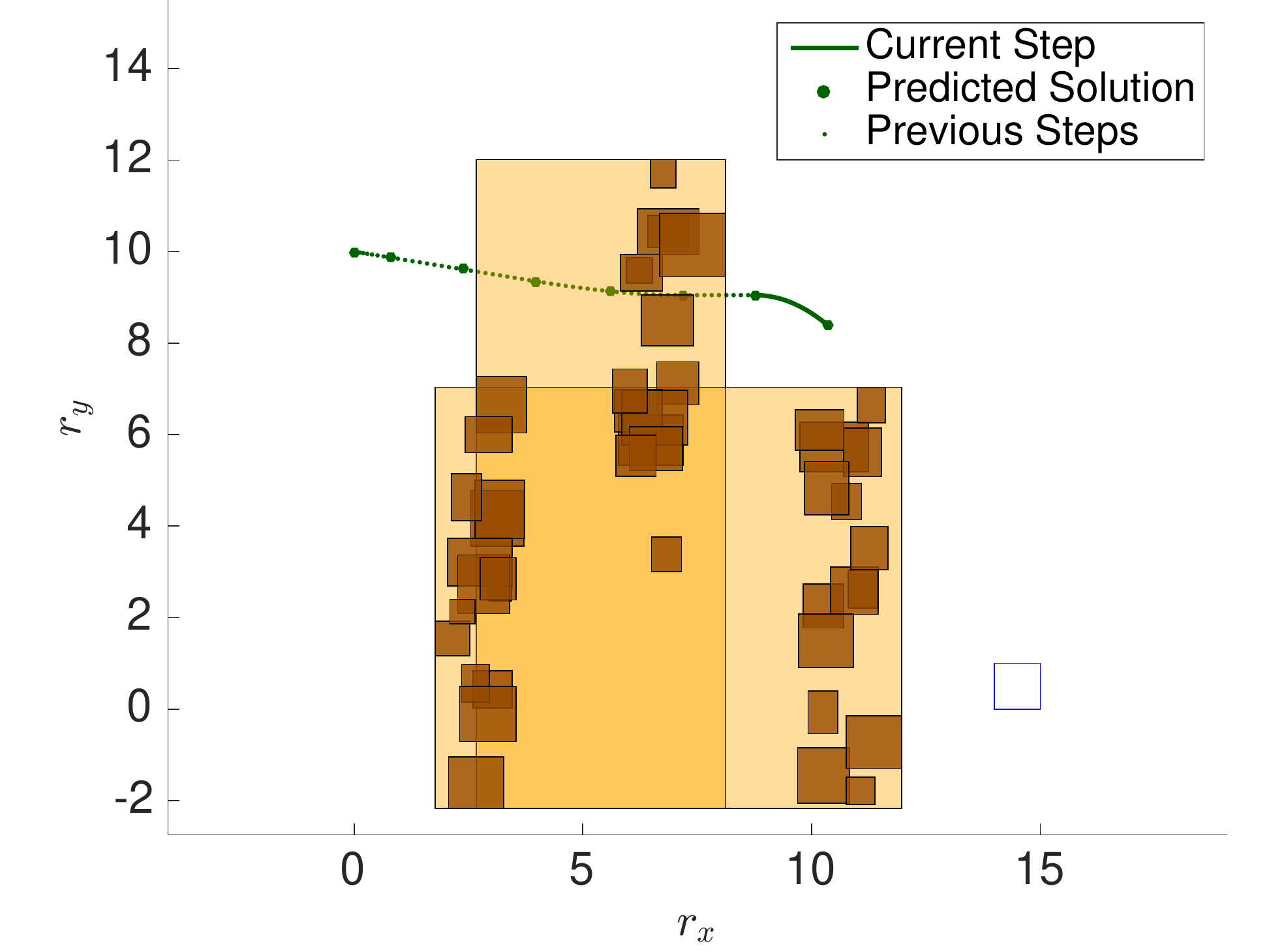} 
		\caption{$k = 7$}
		\label{fig:MILPclust_Nc2_step7}
	\end{subfigure}

	\begin{subfigure}{0.35\textwidth}																																							
		\includegraphics[width=1.00\linewidth]{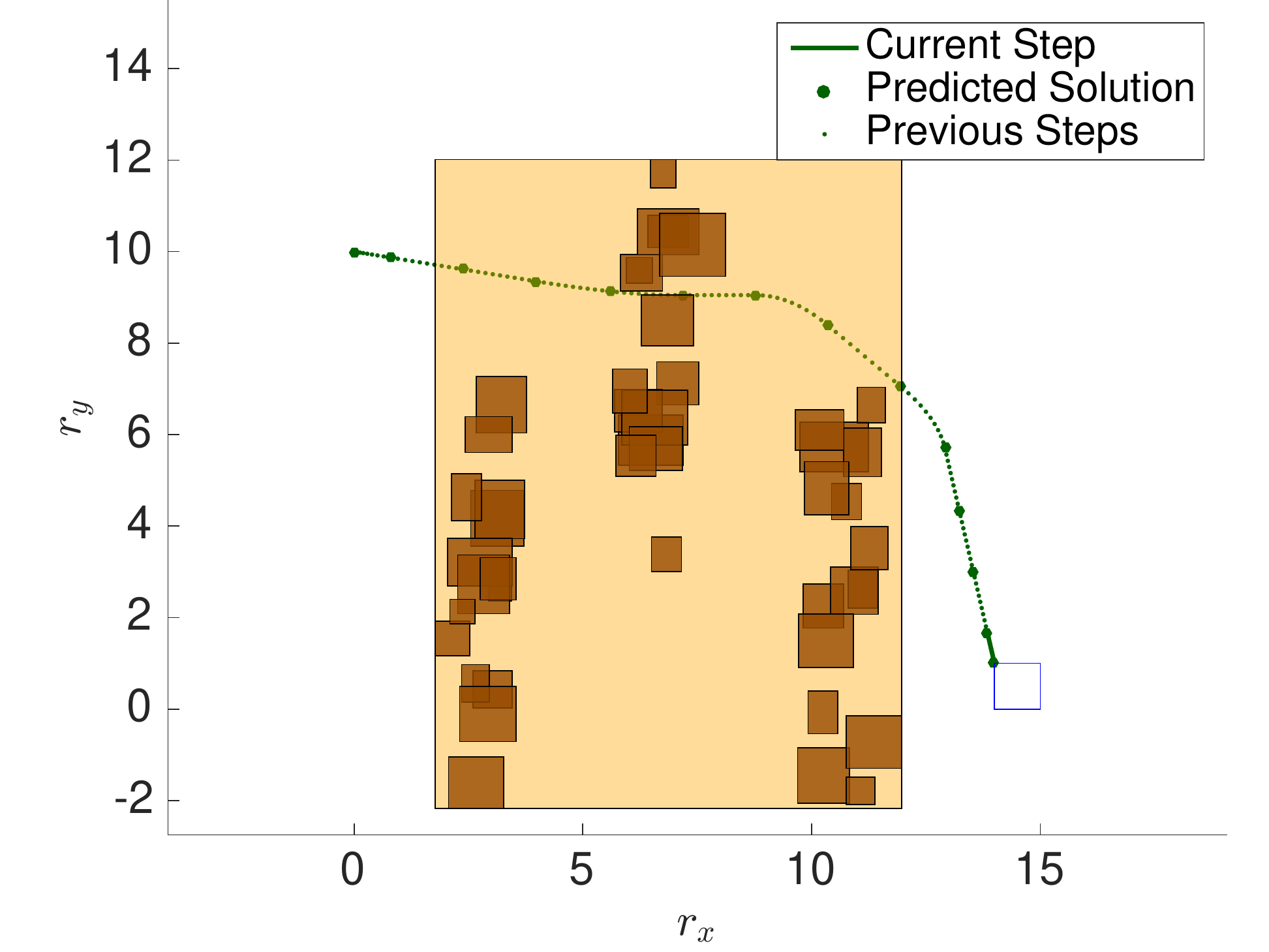} 
		\caption{$k = 13$}
		\label{fig:MILPclust_Nc2_step13}
	\end{subfigure}
\caption{Trajectory and clusters at three different time instants with the clustering strategy with $N_c = 2$.}
\label{fig:MILPclust_Nc2_steps}
\end{figure}

Fig. \ref{fig:MILPclust_Nc2_steps} contains the evolution of the clustering strategy as the agent follows towards the target considering $N_c = 2$ clusters. Initially, in Fig.~\ref{fig:MILPclust_Nc2_step6} the clustering strategy groups $39$ obstacles into a large cluster and the remaining $6$ obstacles into a small cluster, which leaves a small corridor for the agent to proceed towards the target. The agent follows the initially planned path along this corridor until $k = 6$, then the obstacles are assigned differently among clusters, freeing previously clustered space that becomes available for maneuvering. The agent then moves downward, optimizing a shorter path to the target, as can be seen in Fig.~\ref{fig:MILPclust_Nc2_step7}, and after overcoming the obstacles a few steps later is able to reach the target set, clustering the obstacles into a single cluster. 

%
	%
%

\begin{figure}[htpb]
      \centering		
						\includegraphics[width = 0.35\textwidth]{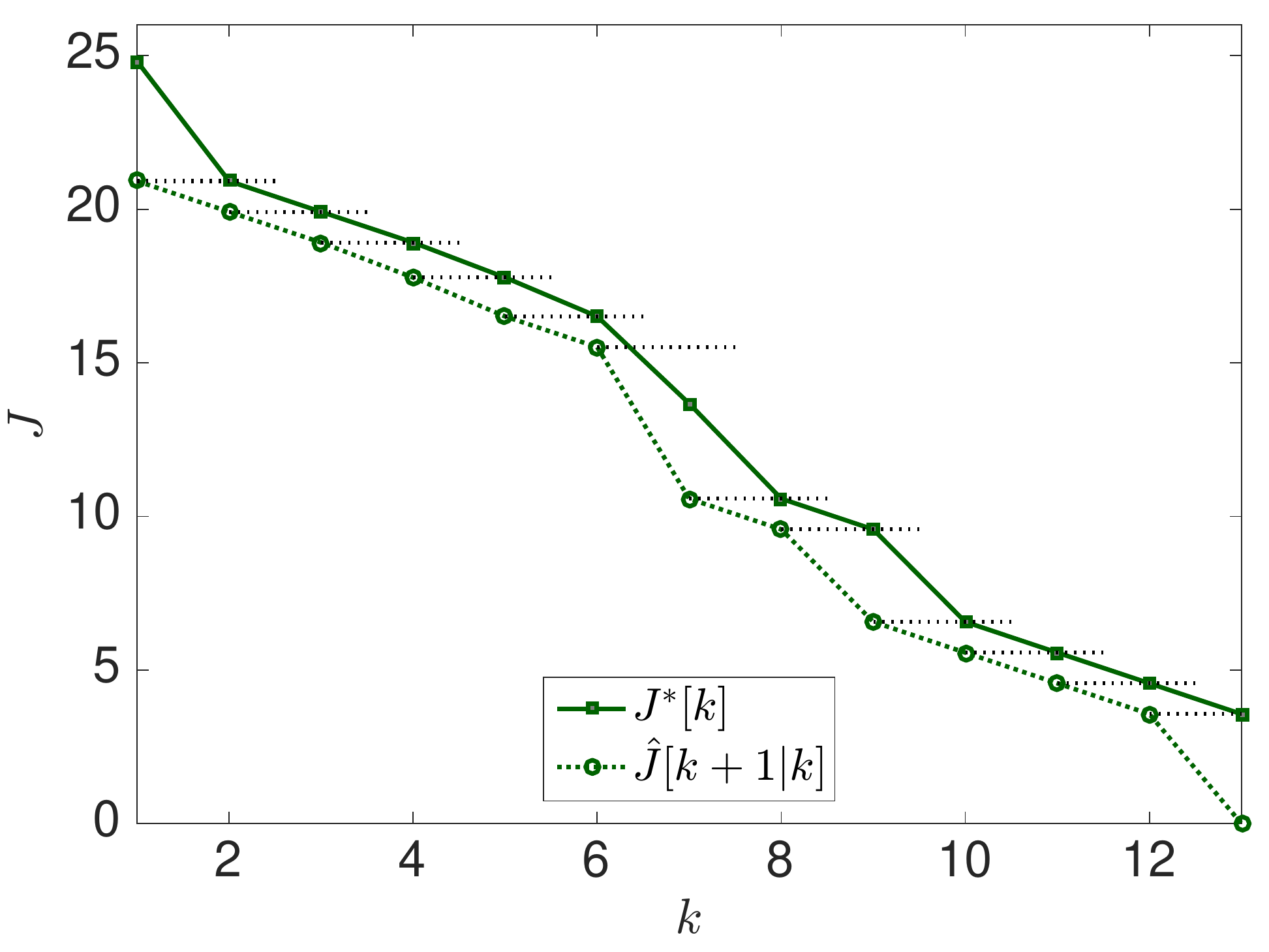}				
      \caption{Evolution of the solution cost and its prediction for the clustering strategy with $N_c=2$ clusters.}
      \label{fig:Jstar_Jhat_milpclust_Nc2}
   \end{figure}

The predicted $\hat{J}$ and optimal $J^{\star}$ costs at each timestep can be seen in Fig. \ref{fig:Jstar_Jhat_milpclust_Nc2}. It can be noticed that $J^{\star}(k+1) = \hat{J}(k+1|k) = J^{\star}(k) - \Vert \textbf{u}(k|k) \Vert_1  - 1, 1 \leq k \leq 6 \ \text{and} \ 7 < k \leq 13$, which coincides with instants where the cluster-obstacle assignment did not change. Otherwise, if a previously clustered space is cleared for maneuvering due to a change in the cluster-obstacle assignment, room for shortcuts not initially considered becomes available, which enables the agent to obtain a shorter path to the target, as can be seen by comparing the clusters in Fig.~\ref{fig:MILPclust_Nc2_step6} and Fig.~\ref{fig:MILPclust_Nc2_step7}. In turn, this yields a lower cost than was predicted, as can be confirmed by $J^{\star}(7) < \hat{J}(7|6)$ in Fig. \ref{fig:Jstar_Jhat_milpclust_Nc2}. For comparison, notice that in Fig.~\ref{fig:Jstar_Jhat_unclustered} the actual optimal cost and the predicted always coincided. Indeed, the final trajectory executed by the agent in closed-loop seen in Fig.~\ref{fig:trajectory_unclustered} is the one predicted at the first time instant, as neither changes in the environment nor disturbances were considered.

As the number of clusters available is limited, the clustering algorithm adjusts the cluster configuration according to the position of the agent to allow a minimum-cost trajectory to the target, which causes the decrease in the value of the cost function regarding what was predicted in the preceding step. 
This illustrates the advantage of our proposal regarding methods that rely on pre-processing the environment, such as the alternatives found in the literature \cite{Vitus2008, Deits2015, Battagello2020}. The embedding of the clustering strategy within the optimization problem enables to decide the clusters considering the minimization of the cost.

\begin{figure}[htbp]
\centering
	\begin{subfigure}{0.35\textwidth}																																						
		\includegraphics[width=1.00\linewidth]{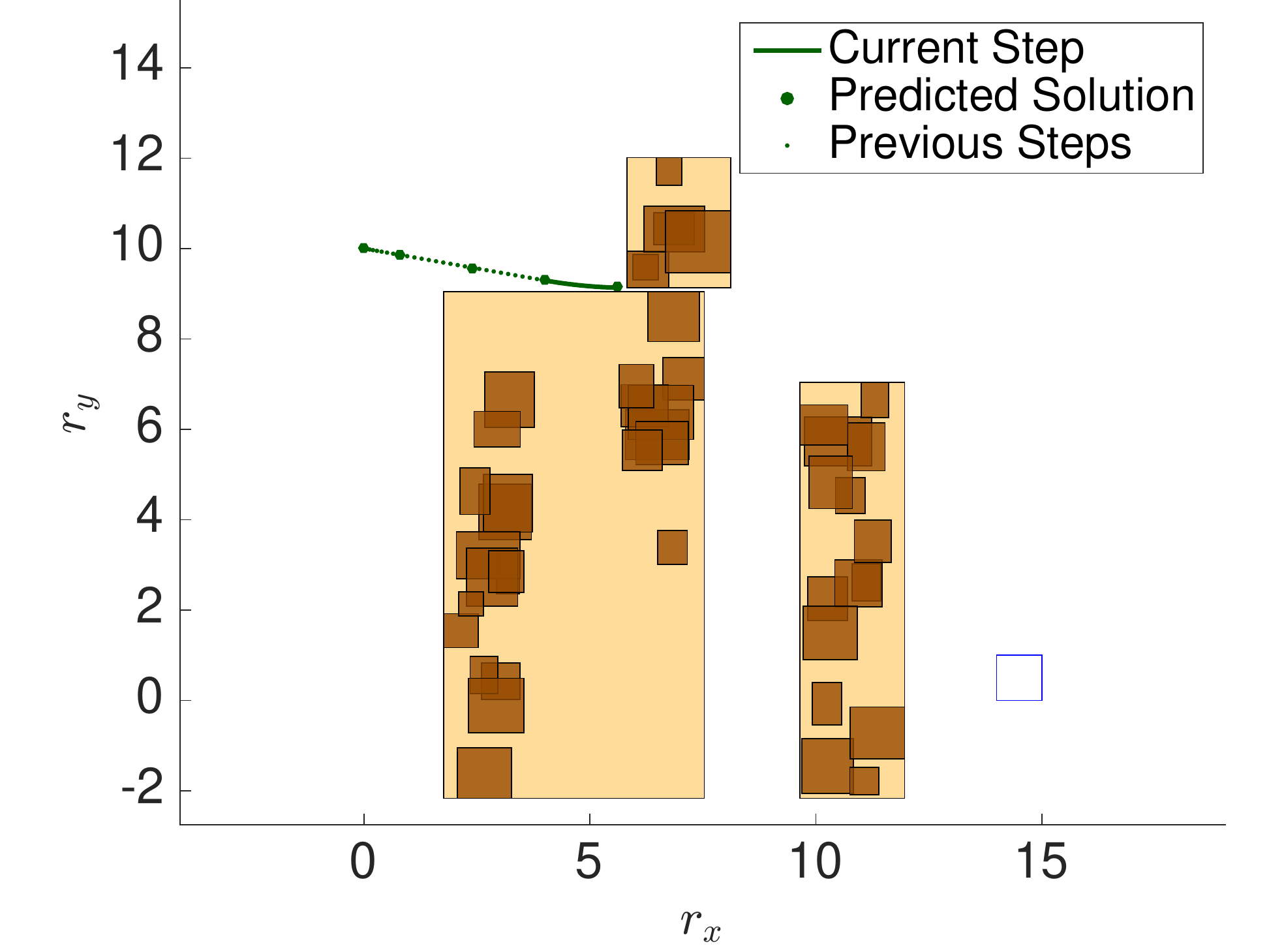}
		
		\caption{$k = 4$}
		\label{fig:MILPclust_Nc3_step4}
	\end{subfigure}
	\begin{subfigure}{0.35\textwidth}																																							
		\includegraphics[width=1.00\linewidth]{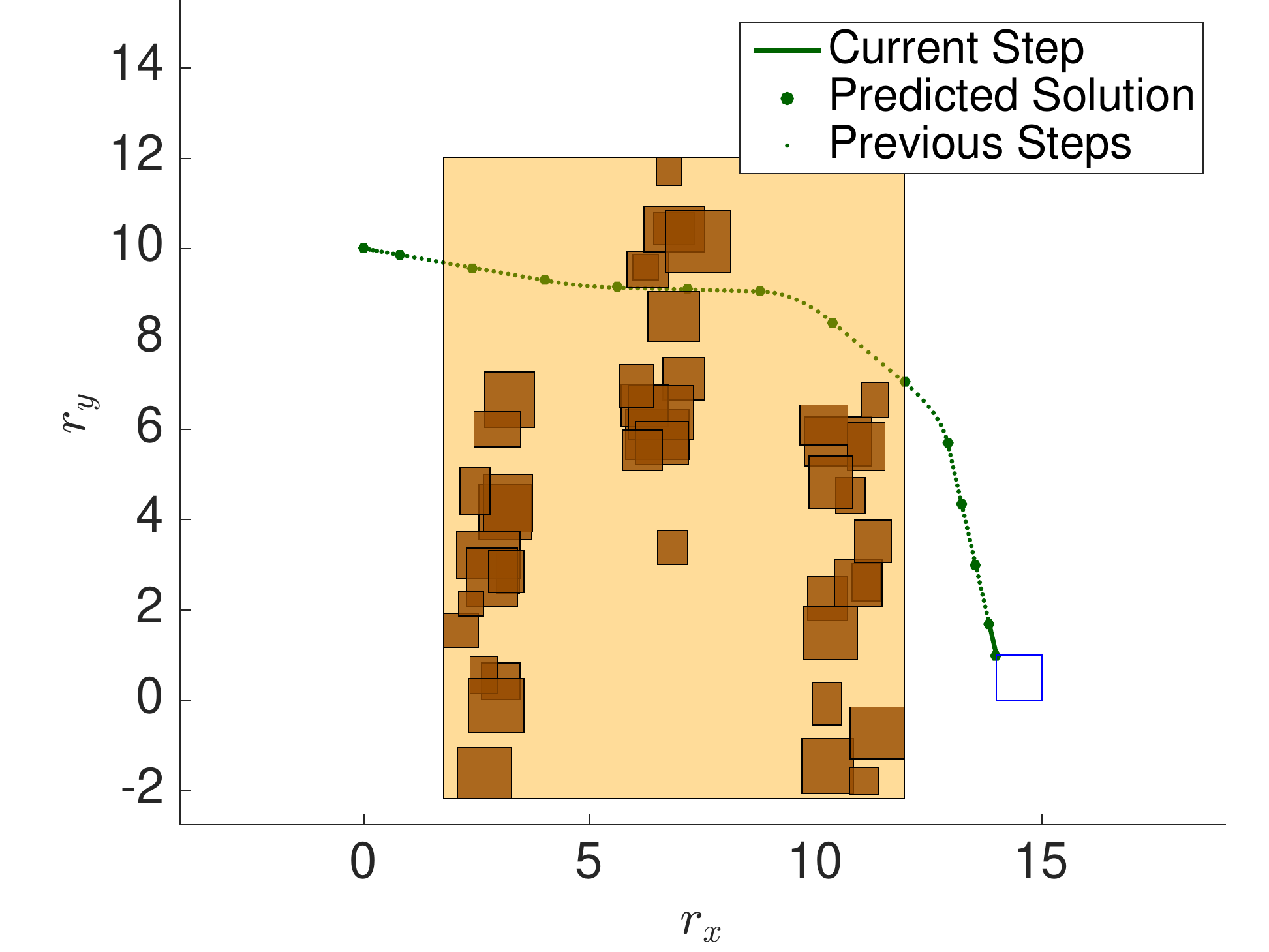}
		
		\caption{$k = 13$}
		\label{fig:MILPclust_Nc3_step13}
	\end{subfigure}
\caption{MILP Clustering strategy with $N_c = 3$.}
\label{fig:MILPclust_Nc3_steps}
\end{figure}

The results for the clustering strategy considering $N_c = 3$ are shown in Fig.~\ref{fig:MILPclust_Nc3_steps}. One can see that the existence of an additional cluster allows to obtain in Fig.~\ref{fig:MILPclust_Nc3_step4} a configuration that already enables a closer circumvention of the rightmost cluster during the last eight steps to reach the target. The previous large cluster of Fig.~\ref{fig:MILPclust_Nc2_step6} is split into two different clusters, one comprising the obstacles in the left hand-side, and another consisting of the remaining obstacles, spreading in the bottom right-hand side of the environment. 
The final trajectory and the last cluster configuration are shown Fig.~\ref{fig:MILPclust_Nc3_step13}.

	%
		%
%
	%
		%

\begin{figure}[htbp]
      \centering			
				\includegraphics[width=0.35\textwidth]{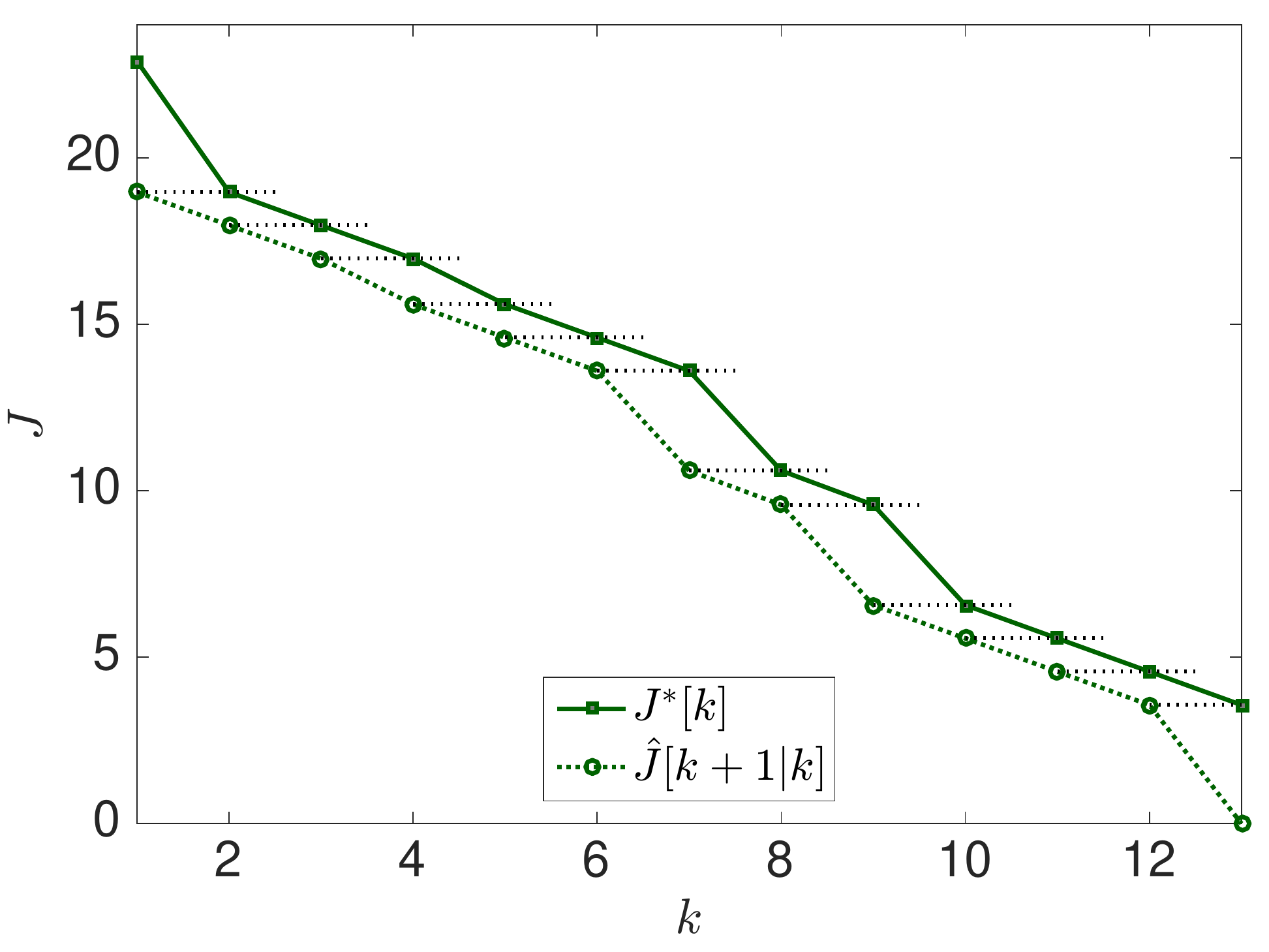}	
      \caption{Evolution of the solution cost and its prediction for the MILP Clustering strategy for $N_c=3$ clusters.}
      \label{fig:Jstar_Jhat_milpclust_Nc3}
   \end{figure}

In Fig. \ref{fig:Jstar_Jhat_milpclust_Nc3} there is a representation of the cost function evolution along the trajectory. 
The clustering procedure obtains clusters for the obstacles such that there is no cost reduction due to changes in the cluster configuration along the agent movement, as opposed to what happened previously for $N_c = 2$.
	
\subsection{Computation time}
In accordance to what was pointed in Sec.~\ref{sec:2-cluster_restrainment}, the MILP Clustering strategy has two main consequences, the first one being its impact upon the optimized trajectory, as discussed in the last section. The second consequence is the reduction of the computation times.
Table \ref{tab:SimulationTimes} depicts the sum of the computation times of every iteration until the vehicle reached the target set and the cost in closed-loop of the final trajectories for the three cases compared in the previous section. In the comparisons that follow, the computation time and the cost of the unclustered strategy will be used as benchmark, corresponding to $100 \%$.  The computation time with $N_c = 2$ is reduced to $5.5\%$ of that obtained without clustering. On the other hand, the cost increases to $106 \%$. Moreover, a compromise can be seen by the clustering with $N_c = 3$, whose computation time is $67\%$ of the unclustered strategy, but the cost is only $102\%$. 
We conjecture that the growth in computation time observed between the scenarios with $N_c = 2$ and $N_c = 3$ clusters is caused by symmetry in the optimization problem. This symmetry is mainly due to different possibilities of assignment of obstacles to clusters that yield the same cost. 

\begin{table}[htbp]
	\centering
		\begin{tabular}{ccc}
		Strategy & Computation time ($s$) & Cost in closed-loop \\
		\hline
		Unclustered & 337.4 & 22.4 \\
		Clustering, $N_c = 2$ & 18.4 & 24.8 \\
		Clustering, $N_c = 3$ & 226.2 & 22.9 			
		\end{tabular}
	\caption{Simulation times in a computer with $2.8$ GHz Intel i7 processor, 8 GB DDR3 RAM, and cost in closed-loop.}
	\label{tab:SimulationTimes}
\end{table}
%
\section{CONCLUSIONS}		\label{sec:3-conclusions}
A novel strategy was proposed in this paper that clusters the obstacles within the optimization problem solved to determine the trajectory. As a result, the choice of the clusters considers directly the optimality of the resulting trajectory. Moreover, this strategy was shown to greatly reduce the computation time of the online solution of the optimization problem, while yielding moderately higher cost. A compromise between the optimality and the computation cost is achieved by the choice of the number of clusters to be used. 

Future work encompasses (i) applying symmetry breaking techniques to reduce the solution search time, particularly when the number of clusters is increased, and (ii) extending the approach to deal with cooperating multiple agents.

\bibliographystyle{IEEEtran}  
\bibliography{references}

\begin{thebibliography}{10}
\providecommand{\url}[1]{#1}
\csname url@rmstyle\endcsname
\providecommand{\newblock}{\relax}
\providecommand{\bibinfo}[2]{#2}
\providecommand\BIBentrySTDinterwordspacing{\spaceskip=0pt\relax}
\providecommand\BIBentryALTinterwordstretchfactor{4}
\providecommand\BIBentryALTinterwordspacing{\spaceskip=\fontdimen2\font plus
\BIBentryALTinterwordstretchfactor\fontdimen3\font minus
  \fontdimen4\font\relax}
\providecommand\BIBforeignlanguage[2]{{%
\expandafter\ifx\csname l@#1\endcsname\relax
\typeout{** WARNING: IEEEtran.bst: No hyphenation pattern has been}%
\typeout{** loaded for the language `#1'. Using the pattern for}%
\typeout{** the default language instead.}%
\else
\language=\csname l@#1\endcsname
\fi
#2}}

\bibitem{Richards2002}
A.~Richards and J.~How, ``Aircraft trajectory planning with collision avoidance
  using mixed integer linear programming,'' in \emph{Proceedings of the 2002
  American Control Conference ({IEEE} Cat. No.{CH}37301)}.\hskip 1em plus 0.5em
  minus 0.4em\relax {IEEE}, 2002.

\bibitem{Richards2006}
A.~Richards and J.~P. How, ``Robust variable horizon model predictive control
  for vehicle maneuvering,'' \emph{International Journal of Robust and
  Nonlinear Control}, vol.~16, no.~7, pp. 333--351, 2006.

\bibitem{Garey1979}
M.~Garey, \emph{Computers and intractability : a guide to the theory of
  NP-completeness}.\hskip 1em plus 0.5em minus 0.4em\relax San Francisco: W.H.
  Freeman, 1979.

\bibitem{Vitus2008}
M.~Vitus, V.~Pradeep, G.~Hoffmann, S.~Waslander, and C.~Tomlin,
  ``Tunnel-{MILP}: Path planning with sequential convex polytopes,'' in
  \emph{{AIAA} Guidance, Navigation and Control Conference and Exhibit}.\hskip
  1em plus 0.5em minus 0.4em\relax American Institute of Aeronautics and
  Astronautics, aug 2008.

\bibitem{Deits2015}
R.~Deits and R.~Tedrake, ``Efficient mixed-integer planning for {UAVs} in
  cluttered environments,'' in \emph{2015 {IEEE} International Conference on
  Robotics and Automation ({ICRA})}.\hskip 1em plus 0.5em minus 0.4em\relax
  {IEEE}, may 2015.

\bibitem{Battagello2020}
V.~A. Battagello, N.~Y. Soma, and R.~J.~M. Afonso, ``Computational load
  reduction of the agent guidance problem using mixed integer programming,''
  \emph{{PLOS} {ONE}}, vol.~15, no.~6, p. e0233441, jun 2020.

\bibitem{Prodan2012}
I.~Prodan, F.~Stoican, S.~Olaru, and S.-I. Niculescu, ``Enhancements on the
  hyperplanes arrangements in mixed-integer programming techniques,''
  \emph{Journal of Optimization Theory and Applications}, vol. 154, no.~2, pp.
  549--572, mar 2012.

\bibitem{Afonso2013}
R.~J.~M. Afonso and R.~K.~H. Galv{\~{a}}o, ``Comments on
  {\textquotedblleft}enhancements on the hyperplanes arrangements in
  mixed-integer programming techniques{\textquotedblright},'' \emph{Journal of
  Optimization Theory and Applications}, vol. 162, no.~3, pp. 996--1003, nov
  2013.

\bibitem{Prodan2015}
I.~Prodan, F.~Stoican, S.~Olaru, and S.-I. Niculescu, \emph{Mixed-Integer
  Representations in Control Design}.\hskip 1em plus 0.5em minus 0.4em\relax
  Springer-Verlag GmbH, 2015.

\bibitem{Bellingham2002}
J.~Bellingham, A.~Richards, and J.~How, ``Receding horizon control of
  autonomous aerial vehicles,'' in \emph{Proceedings of the 2002 American
  Control Conference ({IEEE} Cat. No.{CH}37301)}.\hskip 1em plus 0.5em minus
  0.4em\relax {IEEE}, 2002.

\bibitem{Bemporad1999}
A.~Bemporad and M.~Morari, ``Control of systems integrating logic, dynamics,
  and constraints,'' \emph{Automatica}, vol.~35, no.~3, pp. 407--427, mar 1999.

\bibitem{Franklin1998}
G.~Franklin, \emph{Digital control of dynamic systems}.\hskip 1em plus 0.5em
  minus 0.4em\relax Menlo Park, Calif: Addison-Wesley, 1998.

\end{thebibliography}

	
\end{document}